\documentclass[twocolumn,showpacs,preprintnumbers,amsmath,amssymbaps,pra,citeautoscript]{revtex4}

\usepackage{graphicx}
\usepackage{amssymb}
\usepackage{amsfonts}
\usepackage{subfig}
\usepackage{dcolumn}
\usepackage{bm}
\usepackage{float}
\usepackage{footnote}
\usepackage{dcolumn}
\usepackage{amsmath}
\usepackage{amsfonts}
\usepackage{bm}
\usepackage{xcolor}
\usepackage{epsfig}
\usepackage{amssymb}
\usepackage{latexsym}
\usepackage[colorlinks,linkcolor=blue]{hyperref}
\usepackage{soul}

\newcommand{\be}{\begin{equation}}
\newcommand{\ee}{\end{equation}}
\newcommand{\bea}{\begin{eqnarray}}
\newcommand{\eea}{\end{eqnarray}}

\newcommand{\p}{\partial}
\newcommand{\s}{\sigma}

\newcommand{\la}{\langle}
\newcommand{\ra}{\rangle}
\newcommand{\rd}{\mbox{d}}
\newcommand{\ri}{\mbox{i}}

\newcommand{\lrangle}[1]{\left< #1\right>}

\renewcommand{\vec}[1]{{\bm #1}}

\numberwithin{equation}{section}

\begin{document}
\title{How magnetic field can transform a superconductor into a Bose metal}
\author{Tianhao Ren}
\email{tren@bnl.gov}
\author{Alexei M. Tsvelik}
\email{atsvelik@bnl.gov}
\affiliation{Condensed Matter and Materials Physics Division,  Brookhaven National Laboratory,
  Upton, NY 11973-5000, USA}
 \date{\today } \begin{abstract} We discuss whether a simple theory of superconducting stripes coupled by Josephson tunneling can describe a metallic transport, once the coherent tunneling of pairs is suppressed by the magnetic field. For a clean system, the conclusion we reached is negative: the excitation spectrum of preformed pairs consists of Landau levels, and once the magnetic field exceeds a critical value, the transport becomes insulating. As a speculation, we suggest that a Bose metal can exist in disordered systems provided that the disorder is strong enough to localize some pairs. Then the coupling between propagating and localized pairs broadens the Landau levels, resulting in a metallic conductivity. Our model respects the particle-hole symmetry, which leads to a zero Hall response. And intriguingly, the resulting anomalous metallic state has no Drude peak and the spectral weight of the cyclotron resonance vanishes at low temperatures.
  \end{abstract}

\pacs{74.72.-h, 74.72.Gh, 74.81.-g}

\maketitle

\section{Introduction}

 The discovery of the anomalous superconducting state in the stripe-ordered lanthanum barium copper oxide (LBCO) \cite{TrTs,Tr} has gradually aroused interest in the possible pair density wave (PDW) state - a superconducting state where pairs carry finite momentum (see \cite{review} for the most recent review of the field). At zero field this 3D layered material exhibits an unusual 2D superconductivity with an in-plane Berezinskii-Kosterlitz-Thouless (BKT)  transition coexisting with a finite resistivity along the $c$-axis. This resistivity vanishes at a much smaller temperature, marking an onset of 3D superconductivity with the Meissner effect.  

The recent experiments on several stripe-ordered superconductors have revealed that once the superconductivity is destroyed by the applied magnetic field, a peculiar resistive metallic state with zero Hall response emerges, which persists down to lowest temperatures \cite{tranquada,dragana1,dragana2}. The Hall response vanishes at the BKT temperature, and it remains zero throughout the entire low temperature region even when the superconductivity is destroyed. The critical field is relatively small. The electrical resistance gradually increases with the field, and at fields around $25-30$ T, the sheet resistance reaches a plateau at $R_{\square} \approx 2\pi \hbar/2e^2$. 

Similar anomalous metallic states have been observed in disordered thin films (see, for instance, \cite{PhysRevLett.120.167002} and references therein) and proximity Josephson junction arrays \cite{FigaExp}. The situation with proximity Josephson junction arrays bears the closest resemblance to the one which takes place in the stripe-ordered LBCO, with the difference that the space between the superconducting granules is filled with an ordinary metal \cite{Figa1,Figa2,Figa3}. Since the fermionic quasiparticles are necessary there, one cannot call such state a Bose metal. 

 The purpose of this paper is to find out whether one can describe the anomalous metal without invoking quasiparticles. Experimentally, the stripe-ordered LBCO  appears to be a good candidate for this. Its simplest description \cite{PhysRevLett.88.117001,theory} does not invoke quasiparticles and this is the prism through which we will consider this system. We augment the model by the long-range Coulomb interaction and call it the wire theory (WT for short). WT may be considered as a minimalistic model of PDW since it does not introduce any other entity besides preformed pairs. It describes the stripe-ordered state as an assembly of one-dimensional superconducting wires (stripes) separated by insulating regions and coupled by Josephson tunneling. Each wire constitutes a Luther-Emery liquid, where the interactions generate a spin gap responsible for the superconducting pairing. The bulk superconductivity emerges when the pair tunneling establishes a global coherence. {\it The distinguishing feature} of WT is the suggestion that the Josephson coupling has the wrong sign so that the sign of the superconducting order parameter alternates between the neighboring stripes. Since the stripe orientation changes along the $c$-axis, the sign alternation frustrates the pair tunneling in this direction. It is also assumed that the low energy sector of the system is occupied by bosonic excitations - charged pairs and spin fluctuations from the undoped copper oxide chains between the stripes. If WT is correct and there are no fermionic quasiparticles, then the low temperature resistive state is some kind of Bose metal, where the transport is carried by incoherent pairs.
 
Below we will study WT both in zero and finite magnetic field. We demonstrate that in its simplest form the theory cannot explain the experiments and hence requires certain modifications. It is shown that the excitation spectrum of pairs experiences Landau quantization. Although the nonlinear effects lead to a certain broadening of the Landau levels, in the clean system this effect is not sufficiently strong to lead to metallic transport. As a consequence, once the magnetic field destroys the coherence of pairs, the system becomes an insulator. We suggest as a speculation that a Bose metal can exist in disordered systems provided that the disorder is strong enough to localize some pairs. Due to their small size, these states do not experience Landau quantization and serve as a reservoir of low energy states for the transport. This mechanism allows the existence of a resistive state in the purely bosonic theory of PDW, though it does not explain all the universal features of the transport. An alternative possibility was described in Section III. A. 3 of \cite{KSK}, where one of the authors suggested that the resistive state in the stripe-ordered LBCO owes its existence to the presence of fermionic quasiparticles. This brings us to theories described in \cite{Figa1,Figa2,Figa3}, where superconducting islands embedded in a metallic environment, instead of superconducting stripes coupled by Josephson tunneling, are discussed.

This paper is organized as follows. In Sec. \ref{sec:model}, we introduce the model and discuss its properties in zero magnetic field. In Sec. \ref{sec:RPA}, we discuss the model in finite magnetic field when superconductivity is suppressed. The calculation within the random phase approximation gives us an insulator in a clean system. In Sec. \ref{sec:res}, we discuss the above-mentioned speculation to obtain a Bose metal by introducing localized pairs due to strong disorder. The resulting anomalous metallic state has a zero Hall response due to the particle-hole symmetry, possesses no Drude peak, and has a vanishing cyclotron resonance at low temperatures. These features are consistent with the experiments \cite{tranquada,dragana2,Breznaye1700612,PhysRevLett.120.167002}. Finally in Sec. \ref{sec:discuss}, we make the conclusion. Extra technical details are relegated to the appendices.

\section{The model}
\label{sec:model}
\begin{figure}
	\includegraphics[scale=0.16]{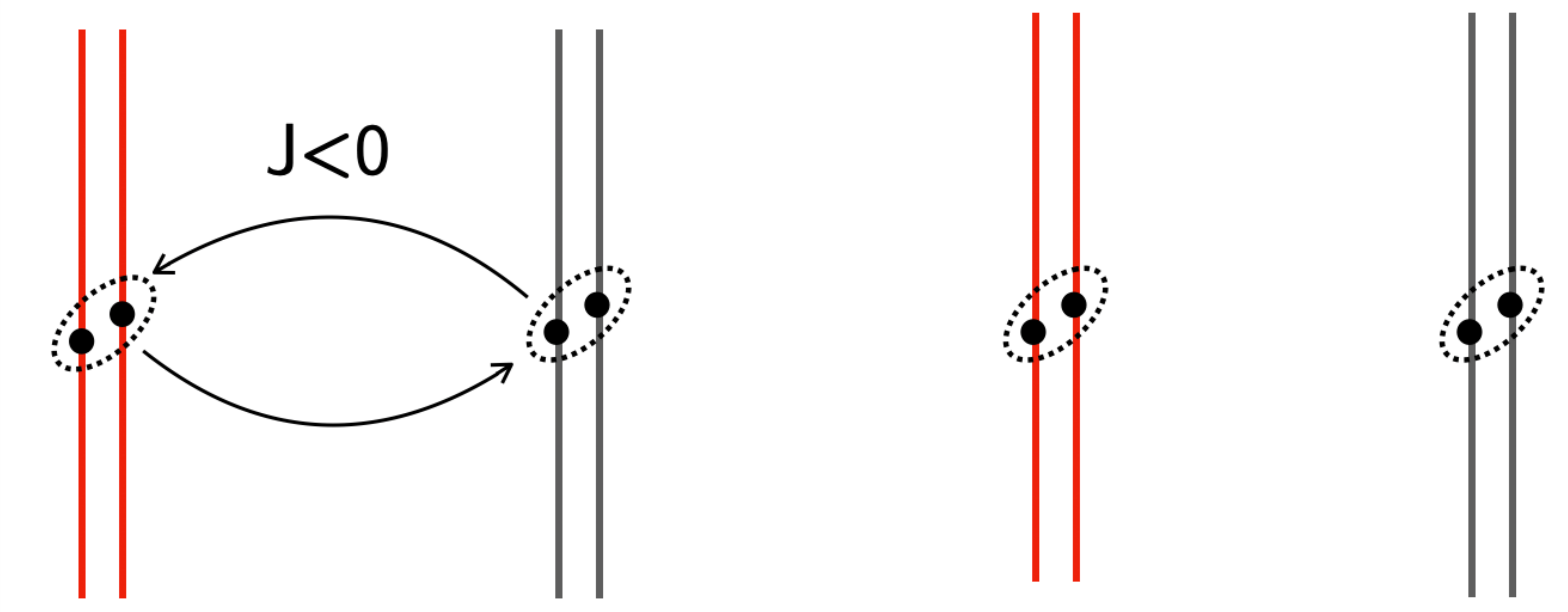}
	\captionsetup{justification=raggedright}
	\caption{The stripe configuration of WT. Each stripe consists of double chains. The Josephson coupling constant between neighboring stripes is negative (shown in the figure only for the first two stripes), such that the superconducting phase fields are pinned at 0 and $\pi$ on alternating stripes (shown in the figure with different colors).}
	\label{fig:double}
\end{figure}
  The model of the stripe-ordered state we consider is of a 3D array of coupled Luther-Emery liquids augmented with the long-range Coulomb interaction, where the $c$-axis coupling is frustrated and can be set to zero. One can consider two versions of it: one where the stripes consist of single doped chains and another where they consist of double chains (see Fig. \ref{fig:double}). In the first (second) case, the superconductivity competes with $2k_F$ $(4k_F)$ charge density wave, or CDW for short, since at weak interactions both susceptibilities are singular. If the paring susceptibility is singular in the second case, the charge density correlations at $2k_F$ are short-ranged (see, for example, \cite{Controzzi}). To simplify matters, we will consider the case when the CDW matrix elements are zero. In this case, the  Lagrangian  can be written as 
\begin{equation}
\label{eq:wire}
	\begin{split}
	L = &\sum_{j} \int \rd x \left[ \frac{v}{2\pi}\left(\p_x\theta_j - \frac{1}{c}A^x_j\right)^2 + \ri \left(\p_{\tau}\theta_j - A^0_j\right)\p_x\phi_j \right.\\ 
		&\left.+ \frac{\pi v}{2}(\p_x\phi_j)^2 - J_{\lrangle{ij}} \cos\left(\theta_i - \theta_j + \frac{1}{c}\int_i^j \rd sA^y(s)\right) \right]  \\ 
		& +\frac{1}{32e^2\pi}\int \rd x\rd y \rd z\left[(\vec\nabla\times \vec A)^2 + \left(\nabla A^0 - \frac{1}{c}\p_{\tau}\vec A\right)^2\right],
	\end{split}
\end{equation}
where a single layer of arrays is considered, since the $c$-axis coupling is frustrated and can be set to zero. Also, we find this form suggested in \cite{aleiner} where dual bosonic fields $\theta,\phi$ are both present more convenient. 

If we adopt the gauge where $\vec\nabla\cdot\vec A =0$,  then the $A^0$ component decouples and can be integrated out. The resulting contribution to the Lagrangian density is 
\bea
\sum_{k \geq j}\int \rd\xi \p_x\phi_j(x)\frac{4e^2}{[\xi^2 + a_0^2(k-j)^2]^{1/2}}\p_x\phi_k(x+\xi),
\eea
where $a_0$ is the distance between the stripes. Since the backscattering term is absent in the current case, we can also integrate over $\phi$ fields and obtain a closed expression for the Lagrangian:
\begin{equation}
	\begin{split}
		L = & \sum_{i,j}\int \rd x \left[\frac{1}{2}\p_{\tau}\theta_i C^{-1}_{ij}\p_{\tau}\theta_j + \frac{v}{2\pi}\left(\p_x\theta_j - \frac{1}{c}A^x_j\right)^2\right.\\
		&\left. - J_{\lrangle{ij}} \cos\left(\theta_i - \theta_j + \frac{1}{c}\int_i^j \rd sA^y(s)\right) \right] \nonumber \\
		& + \frac{1}{32e^2\pi}\int \rd x\rd y\rd z (\vec\nabla\times \vec A)^2,
	\end{split}	
\end{equation}
where the matrix $C_{ij}$ is defined through
\begin{equation}
	\begin{split}
	&\sum_{i,j}\partial_x\phi_i C_{ij}\partial_x\phi_j=\sum_i\pi v\left(\partial_x\phi_i\right)^2\\
	&+\sum_{i,j}\int \rd\xi \p_x\phi_i(x)\frac{4e^2}{[\xi^2 + a_0^2(i-j)^2]^{1/2}}\p_x\phi_j(x+\xi).
	\end{split}
\end{equation}
This completes our construction of WT. Before we consider the transport properties of WT, we first discuss two relevant issues: the plasmon mode in zero magnetic field and the pairing susceptibility of a single stripe.

\subsection{The plasmon mode in zero magnetic field}

 In the superconducting phase where $\theta$ fields are pinned either at 0 ($J >0$) or at 0 and $\pi$ on alternating stripes ($J <0$). The later case can be reduced to the former by the substitution $\theta_j = j\pi  + \tilde\theta_j$ where $\tilde\theta$ is a slow function of coordinates. Then we can expand the cosine term and obtain the spectrum
\bea
	\omega^2 = C(q)\left[\pi^{-1} v^2 q_x^2 + 4|J|\sin^2(q_y a_0/2)\right], \label{plasma2}
\eea
where $a_0$ is the distance between the stripes and $C(q)$ is the Fourier component of the matrix $C_{ij}$. In the arrangement relevant to stripe-ordered cuprates, the stripes comprise a three dimensional structure and the Fourier transform of $C_{ij}$ depends on all three components of the wave vector. At small wave vectors $C(q) \sim |{\bf q}|^{-2}$, therefore the spectrum in 3D is gapped, as is customary for 3D  plasmons. In literature related to the stripe-ordered states, the Goldstone excitations in  Fulde-Ferrel-Larkin-Ovhinnikov (FFLO) superconductor have been considered \cite{samokhin2010}, but the issue of the Coulomb interaction were not addressed.

As for experimental data, the value of the in-plane plasma frequency for $x=1/8$ LBCO extracted from the optical measurements \cite{Homes2012} is around $1600 \text{ cm}^{-1}$ (200 meV). The same experiments give the  spectral gap the estimate around $20$ meV so that the plasma frequency is well above this cutoff, and we can ignore the effect of the plasmon mode in the following discussions. 

\subsection{Pairing susceptibility for a single stripe}
\label{sec:pair}

 As far as pairing susceptibility is concerned, we will show that the long-range nature of the Coulomb interaction does not change matters qualitatively. For this, we calculate the pairing susceptibility for a single stripe. We assume that the system is 3D and integrate over the transverse momentum $Q$ over the 2D slice of the Brillouin zone. In what follows, we neglect the $q_z$-dependence of $C(Q,q_z)$. Then the pairing susceptibility for a single stripe is obtained as
\begin{equation}
	\begin{split}
		& \chi_P(\tau,x)\\
		\sim & \exp\left[ -\mathcal{N}\int \frac{\rd^2 Q}{(2\pi)^2}\int\frac{\rd\omega\rd q}{(2\pi)^2}\frac{1 - e^{-\ri\omega\tau +\ri qx}}{\omega^2/vC(Q) + vq^2/\pi}\right] \\
		=& \exp\left[-\frac{\mathcal{N}}{4 v}\int \frac{\rd^2 Q v(Q)}{(2\pi)^2}\ln\left(\frac{v^2(Q)\tau^2 +x^2}{a_0^2}\right)\right],
	\end{split} 
\end{equation}
where $v(Q) = v[C(Q)/\pi]^{1/2}$ and $\mathcal{N}$ is a normalization factor such that we get back to the noninteracting result when we set $e=0$. This function can be approximately replaced by the standard power law with the scaling dimension 
\be
d = d_0\Big[\sqrt{1+\alpha} + \alpha \ln(\alpha^{-1/2} + \sqrt{1 + \alpha^{-1/2}})\Big],
\ee
where $\alpha = 16e^2/\pi, d_0=1/4$. As a result, we can imitate the effect of the long-range Coulomb interaction as equivalent to the renormalization of the Luttinger parameter. In what follows, we focus on the case $d\ll 1$ when the Josephson tunneling is relevant.

\section{Finite magnetic field. RPA approach}
\label{sec:RPA}

In this section we study the low temperature regime in strong magnetic field when superconductivity is suppressed. We treat the magnetic field inside the sample as uniform. We expect this approximation to be valid in strong magnetic fields. Below we consider the case with $J<0$. In our analysis, we use the random phase approximation (RPA), which replaces the original action for the order parameter field $\Phi=\exp(\ri \tilde{\theta})$ by the Gaussian one:
\begin{equation}
	\begin{split}
	S = &\sum_{\omega,q,j}\Phi_j^+(\omega,q)\chi_P^{-1}(\omega,q)\Phi_j(\omega,q) \\
	& - J\sum_{\omega,j}\int \rd x \Big[\Phi^+_j(\omega,x)e^{2\ri eHa_0x/c}\Phi_{j+1}(\omega,x) + H.c.\Big],
	\end{split}
\end{equation}
where $\chi_P$ is the pairing susceptibility for a single stripe calculated in Section \ref{sec:pair}. The formal expansion parameter of RPA is the inverse number of nearest neighbors, so we expect our results to be valid only qualitatively. In the momentum representation things become even more convenient:
\begin{equation}
	\begin{split}
	S &= \sum_{\omega,q,p}\Phi^+(\omega,q;p)
	\Big[\chi_P^{-1}(\omega,q)\Phi(\omega,q;p)\\
	& - Je^{\ri p a_0}\Phi(\omega,q-h;p) - Je^{-\ri pa_0}\Phi(\omega,q+h;p)\Big],
	\end{split}
\end{equation}
where $h = 2eHa_0/c$. To be compared with the experimental data, it is convenient to express it as
\be 
ha_0 = \frac{2\mu_BH}{\hbar^2/2m_ea_0^2}, \label{h}
\ee
where $\mu_B$ is the Bohr magneton and $m_e$ is the electron mass. To calculate the Green's function $G\equiv -\ri\la \Phi\Phi^{\dagger}\ra$, we need to solve the equation 
 \begin{equation}
 \label{eq:GG}
 	\begin{split}
 		\Big[\chi_P^{-1}&(\omega,q)\delta_{q,q'} -Je^{\ri pa_0}\delta_{q,q'+h} \\
 		&  - Je^{-\ri pa_0}\delta_{q,q'-h}\Big] G(\omega,q',q'';p) = \delta_{q,q''}.
 	\end{split}	
 \end{equation}
The formal solution is expressed via normalizable eigenfunctions satisfying 
\be
\Big[\chi_P^{-1}(\omega,q)\delta_{q,q'} - J\delta_{q,q'+h} - J\delta_{q,q'-h}\Big]\Psi(q') = \lambda\Psi(q), \label{eigen1}
\ee
such that the Green's function can be expressed as
\be
\label{eq:disfree}
 G(\omega, q,q';p) = \sum_{\lambda}e^{\ri pa_0(q-q')/h}\frac{\Psi_{\lambda}(q)\Psi^+_{\lambda}(q')}{\lambda}.
\ee

Taking into account that $a_0 \approx 1.5$ nm we have $\hbar^2/(4m_ea_0^2) \approx 10^2$ K. For $H = 30$ T, we obtain $ha_0 \approx 1/3$, which means that for the entire experimental range of \cite{tranquada,dragana1,dragana2}  it is reasonable to adopt  $ha_0 \ll 1$. Then  we can reformulate the eigenvalue problem (\ref{eigen1}) as a differential equation:
\be
\Big[\chi_P^{-1}(\omega,q) - 2J\Big]\Psi(q) - Jh^2\frac{\rd^2}{\rd q^2}\Psi(q) =\lambda\Psi(q). \label{eigen}
\ee
It is more convenient to perform the calculations in imaginary time, since then the Green's function is real, and to do analytic continuation afterwards. 
For $T=0$ we have
\bea
\chi_P^{-1}(\ri\omega_n,q) = -(\omega_n^2 +v^2q^2)^{1-d}\Lambda^{2d}.
\eea
where $\Lambda$ is the high energy cutoff, and we will set $\Lambda =1$ and restore it when necessary. Let us perform the rescaling 
\be
\omega_n = \epsilon_0\Omega_n, ~~ vq = \epsilon_0 x, ~~ \epsilon_0 = (|J|h^2v^2)^{\frac{1}{2(2-d)}},
\ee
such that Eq. (\ref{eigen}) assumes the dimensionless form:
\begin{equation}
	\begin{split}
		&\Big[(x^2 + \Omega_n^2)^{1-d}  - \frac{\rd^2}{\rd x^2}\Big]\Psi_k(x) =f_k(\ri\Omega_n )\Psi_k(x), \label{Sch}\\
		& \lambda = 2|J| - (|J|h^2v^2)^{(\frac{1-d}{2-d})}f_k, ~~ k=0,1,\cdots,
	\end{split}
\end{equation}
where the eigenfunctions $\Psi_k(x)$ can be chosen to be real. Then the Green's function after analytic continuation $\ri\omega_n\to \omega+\ri 0$ becomes 
\begin{equation}
	\begin{split}
		& G^R(\omega, q,q';p)\\
		=&\sum_{k=0}^{\infty}e^{\ri pa_0(q-q')/h}\frac{\Psi_{k}(vq/\epsilon_0)\Psi_{k}(vq'/\epsilon_0)}{2|J|-(|J|h^2v^2)^{\frac{1-d}{2-d}}f_k(\omega/\epsilon_0)}.
	\end{split}	
\end{equation}
The pole of this Green's function at $\omega=0$ marks the critical magnetic field above which superconductivity will be lost. Since the eigenvalue in Eq. (\ref{Sch}) for $\Omega_n =0$ is on the order of one, the critical magnetic field $h_c$ can be determined as
\be
\label{eq:renormH}
vh_c \sim \Lambda \left(\frac{|J|}{\Lambda^2}\right)^{\frac{1}{2(1-d)}}, ~~ \epsilon_0^{c} = \Lambda\left(\frac{2|J|}{f_0(0)\Lambda^2}\right)^{\frac{1}{2(1-d)}}.
\ee
It should be noted that the present calculation is valid only for $h>h_c$, when the cyclotron radius for pairs $v/\epsilon_0$ is finite. 

To simplify the expressions, below we will set $\epsilon_0=v=1$, and restore them when necessary. For real $\Omega_n$, Schr\"odinger equation (\ref{Sch}) has a discrete spectrum with eigenvalues depending on $\Omega_n$. The wave functions for even $k$ are even, for odd $k$ are odd and vanish at $x=0$. It is interesting to express the Green's function in real space:
\begin{equation}
	\begin{split}
		G^R(\omega,& x_2,x_1,y) = \frac{e^{-\frac{\ri h y(x_1+x_2)}{2a_0}}}{2J} \sum_{k=0}^{\infty}\int \frac{\rd q}{(2\pi)^2} \\
		&\times e^{\ri q(x_2-x_1)}\frac{\Psi_k\left(q-\frac{yh}{2a_0}\right)\Psi_k\left(q+\frac{yh}{2a_0}\right)}{(h/h_c)^{\frac{(2-2d)}{(2-d)}}f_k(\omega)/f_0(0)-1}.
	\end{split}
\end{equation}
It is obvious that for $d \neq 0$, functions $f_k(\ri\Omega_n)$ have branch cuts so that under analytic continuation $\ri\Omega_n \rightarrow \omega +\ri 0$ they become complex. This means that the discrete levels of the Schr\"odinger equation (\ref{Sch}) acquire finite width. Moreover, since we are considering stripes made of double chains, CDW is suppressed and the CDW operator will not couple strongly to disorder.  

We can calculate $\delta f_k(\omega)=f_k(\omega)-f_k(0)$ for small $\omega$ using the first order perturbation theory:
\be
  	\delta f_k(\ri\Omega_n) = \int \rd x \Psi^2_k(x)\Big[(x^2 +\Omega_n^2)^{1-d} - |x|^{2(1-d)}\Big],
\ee
followed by the analytic continuation $\ri\Omega_n\to \omega+\ri 0$. Then the imaginary and real parts of $\delta f_k(\omega)$ are
\begin{equation}
	\begin{split}
		& \Im m ~\delta f_k=-2\sin(\pi d)\operatorname{sgn}\omega\int_0^{\left|\omega\right|}\left[\omega^2 -x^2\right]^{1-d}\Psi_k^2(x)\rd x, \\
		& \Re e~\delta f_k \approx - (1-d)\omega^2\int \rd x \Psi^2_k(x)x^{-2d}\rd x.
	\end{split}
\end{equation}
The calculation of the imaginary part depends on $k$. For even $k$, we have
\begin{equation}
	\Im m ~\delta f_k\approx -2\sin(\pi d)\operatorname{sgn}\omega\Psi_k^2(0)\int_0^1(1-x^2)^{1-d}\rd x~|\omega|^{3-2d},
\end{equation}
while for odd $k$, we have
\begin{equation}
	\Im m ~\delta f_k\approx -\sin(\pi d)\operatorname{sgn}\omega\int^1_0(1-x^2)x^2dx~|\omega|^{5-2d}.
\end{equation}
To summarize, we use the following short notations:
\begin{equation}
\label{Im}
	\begin{split}
	& \Im m ~\delta f_k \approx  \begin{cases}
 	-\alpha_k(d)|\omega|^{3-2d}\operatorname{sgn}\omega & k \text{ is even} \\
 	-\alpha_k(d)|\omega|^{5-2d}\operatorname{sgn}\omega & k \text{ is odd}
 	\end{cases},\\
	& \Re e ~\delta f_k  = - \beta_{k}(d)\omega^2,
	\end{split}
\end{equation}
where $\alpha_k, \beta_k$ are dimensionless constants depending on the scaling dimension $d$. Close to the critical field at frequencies $\omega\ll \epsilon_0$ we can write the Green's function as 
\begin{equation}
\label{eq:disorderfree}
	G^R(\omega, q,q';p) = \frac{e^{\ri pa_0(q-q')/h}}{2J}\sum_{k=0}^{\infty}\Psi_k(q)\Psi_k(q')\chi_k(\omega),
\end{equation}
where the first term in the summation is
\begin{equation}
\label{eq:chi0}
	\chi_0(\omega)=\frac{1}{\left[(h/h_c)^{\frac{(2-2d)}{(2-d)}} -1\right] - \beta_0\omega^2  - \ri \alpha_0|\omega|^{3-2d}\operatorname{sgn}\omega},
\end{equation}
and the other terms with nonzero $k$ are similar. It can be seen from the poles of $G^R$ that the excitation spectrum of the pairs corresponds to damped Landau levels. Another important feature of the Green's function is that it possesses the particle-hole symmetry such that $G^R(\omega)=[G^R(-\omega)]^*=G^A(-\omega)$. This will lead to a zero Hall response, which we will discuss in Sec. \ref{sec:conduct}.

\subsection{Conductivities}

\label{sec:conduct}

The current operators for currents along and transverse to the stripe direction are given by the following expressions:
\be
J_x = \ri\Phi^+_j\p_x\Phi_j, ~~ J_y = \ri J\left(\Phi^+_j\Phi_{j+1}e^{\ri hx} - H.c.\right).
\ee
The complete Kubo formula for the conductivity is
\begin{equation}
	\sigma_{\alpha \beta}(\boldsymbol{q}, \omega)=\frac{1}{\omega} \int_{0}^{\infty} \rd t e^{\ri \omega t}\left\langle\left[J_{\alpha}^{\dagger}(\boldsymbol{q}, t), J_{\beta}(\boldsymbol{q}, 0)\right]\right\rangle+\ri\frac{vK}{\pi\omega}\delta_{\alpha\beta},
\end{equation}
where $K=d_0/d$ is the Luttinger parameter. The second term in $\sigma_{\alpha\beta}(\bm{q},\omega)$ is the diamagnetic term, and it will be cancelled by a corresponding contribution from the first term, since we are considering the case $h>h_c$ where superconductivity is lost. Consequently, only the real part of the conductivity remains:
\begin{equation}
\label{eq:Kubo}
	\operatorname{Re} \sigma_{\alpha \beta}(\boldsymbol{q}, \omega)=\frac{1-e^{-\beta \omega}}{2 \omega } \int_{-\infty}^{\infty} \rd t e^{\ri \omega t}\left\langle J_{\alpha}^{\dagger}(\boldsymbol{q}, t) J_{\beta}(\boldsymbol{q}, 0)\right\rangle.
\end{equation}
This form of Kubo formula can be understood as a kind of fluctuation-dissipation theorem. We now focus on the DC conductivity at $h>h_c$, which can be calculated by setting $q=0$ in Eq. (\ref{eq:Kubo}) and taking the limit $\omega\to 0$. Firstly, the conductivity along the stripe direction is obtained as 
\begin{equation}
\label{eq:sigmaxx}
	\begin{split}
		&\s_{xx}(\omega)= \sum_{p,k_{1},k_2}\left(\int \rd q q\Psi_{k_1}(q) \Psi_{k_2}(q)\right)^2 F_{k_1,k_2}(\omega),\\
		& F_{k_1,k_2}(\omega)=\frac{\ri}{\omega}\Pi_{k_1,k_2}(\omega),\\
		& \Pi_{k_1,k_2}(\ri \Omega_n) = \frac{T}{4J^2}\sum_{\omega_m}\chi_{k_1}(\ri\omega_m)\chi_{k_2}(\ri\Omega_n + \ri\omega_m).
	\end{split}
\end{equation}
As usual, the summation over $p$ gives $\Phi/\Phi_0$, where $\Phi=HS,\Phi_0=2\pi\hbar c/(2e)$. The matrix elements in Eq. (\ref{eq:sigmaxx}) are nonzero only when the eigenfunctions $\Psi_k$ have different parity. The frequency sum gives 
\begin{equation}
	\begin{split}
		\Pi_{k_1,k_2}(\ri\Omega_n) = \frac{1}{4J^2\pi^2 }\int &\rd \xi\rd \xi' \frac{\coth(\xi/2T) - \coth(\xi'/2T)}{\ri \Omega_n +\xi -\xi'}\\
		&\times\Im m \chi_{k_1}(\xi)\Im m \chi_{k_2}(\xi').
	\end{split}
\end{equation}
Performing the analytic continuation $\ri\Omega_n\to \omega+\ri0$ and we obtain
\begin{equation}
\label{eq:F12}
	\begin{split}
		F_{k_1,k_2}(\omega) = \frac{1}{4J^2\pi }\int & \rd y \frac{\coth(y/2T) - \coth[(y+\omega) /2T]}{\omega}\\
		&\times \Im m \chi_{k_1}(y)\Im m \chi_{k_2}(y+\omega).
	\end{split}
\end{equation}
If we perform a shift of the integration variable $y\to y-\omega$ followed by a reflection $y\to -y$, using the particle-hole symmetry that $\Im m \chi_k(\omega)=-\Im m\chi_k(-\omega)$, we then obtain $F_{k_1,k_2}(\omega)=F_{k_2,k_1}(\omega)$. This ensures that the summation over $k_1$ and $k_2$ won't make the longitudinal conductivity $\sigma_{xx}(\omega)$ vanish.

For magnetic fields above the transition point in the limit of zero frequency we obtain
\begin{equation}
\label{eq:conxx}
		\s_{xx}(\omega\to0) \sim \int \rd y \frac{|y|^{8-4d}}{2T\sinh^2(y/2T)} \sim \s_0(T/\epsilon_0)^{8-4d},
\end{equation}
where it is assumed that the scaling dimension $d\ll 1$. We can see that the DC conductivity $\s_{xx}(\omega\to 0)$ decreases with temperature, resulting in an insulator, though without a sharp gap.

Next we consider the conductivity transverse to the stripe direction. It is obtained as
\begin{equation}
	\begin{split}
		\sigma_{yy}(\omega)=J^2&\sum_{p,k_1,k_2}\bigg{\{}\int dq\Psi_{k_1}(q)\left[\Psi_{k_2}(q-h)\right. \\
		&\left.-\Psi_{k_2}(q+h)\right] \bigg{\}}^2F_{k_1,k_2}(\omega).
	\end{split}
\end{equation}
This expression is even in $h$ as it should be, and it has a similar structure as $\sigma_{xx}(\omega)$, where the matrix elements are nonzero only when the eigenfunctions $\Psi_{k}$ have different parity. Consequently, $\sigma_{yy}(\omega\to 0)$ has the same behavior as $\sigma_{xx}(\omega\to 0)$ at low temperatures, vanishing with the same power as shown in Eq. (\ref{eq:conxx}). 

Finally, we consider the Hall conductivity, which is obtained as
\begin{equation}
\label{eq:Hall}
	\begin{split}
		\sigma_{xy}&(\omega)=-\ri J\sum_{p,k_1,k_2}\bigg{\{}\int \rd q \Psi_{k_1}(q)\left[\Psi_{k_2}(q-h)\right.\\
		 &\left.-\Psi_{k_2}(q+h)\right]\bigg{\}}\cdot \left( \int \rd q q\Psi_{k_2}(q)\Psi_{k_1}(q) \right)F_{k_1,k_2}(\omega).
	\end{split}
\end{equation}
This expression is odd in $h$ as it should be. Since $F_{k_1,k_2}(\omega)=F_{k_2,k_1}(\omega)$ as shown above, the summand in this expression is odd under the exchange $k_1\leftrightarrow k_2$. As a result, the summation over $k_1$ and $k_2$ makes $\sigma_{xy}(\omega)$ vanish identically. This shows that the particle-hole symmetry ensures a zero Hall conductivity, which is consistent with the experiments \cite{tranquada,dragana2,Breznaye1700612}.

To summarize, the DC transport calculation in this section shows that both longitudinal conductivities, along or transverse to the stripe direction, vanishes as powers of $T$ at low temperatures, while the Hall conductivity vanishes identically. As a result, when superconductivity is suppressed, WT gives out an insulator with a soft gap on the level of RPA, excluding the possibility of a Bose metal in a clean system.

\section{A Possible Escape Route}
\label{sec:res}

\begin{figure}
	\includegraphics[scale=0.16]{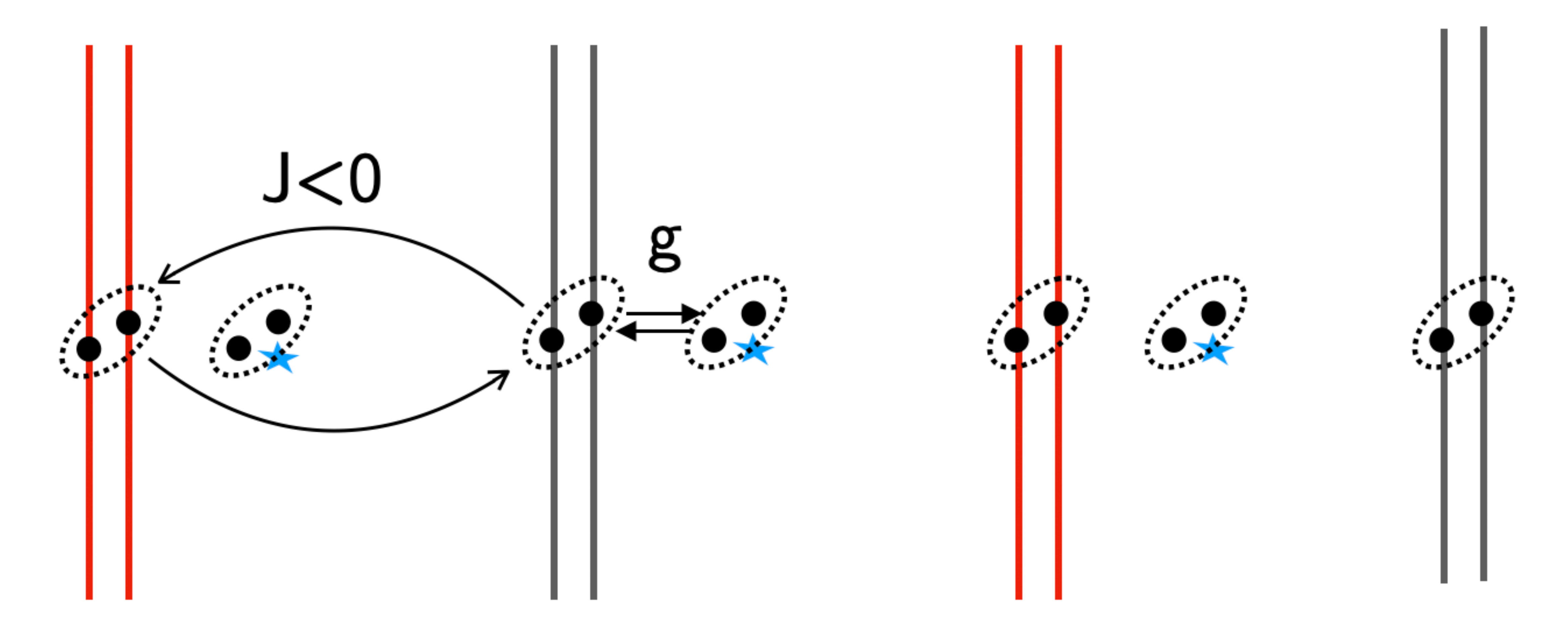}
	\captionsetup{justification=raggedright}
	\caption{Localized pairs in between the stripes, where the star in the figure denotes the disorder. The coupling between the order parameter field and the localized pair is shown in the figure only for the second stripe.}
	\label{fig:disorder}
\end{figure}

So it appears that the simple WT in a clean system is not capable to describe the metallic state. A possible mechanism to produce a metallic state on top of WT is to imagine that the system supports localized pairs due to strong disorder. Here we just suggest this mechanism as a possibility without providing any microscopic justification for it.

Our speculation comes from the suggestion that above some critical magnetic field $\sim h_c$, there is an intermediate region between the superconducting stripes, termed as charged insulator \cite{Ren} . This region can accommodate charges in the form of pairs localized by quenched disorder (see Fig. \ref{fig:disorder}). Due to the small size of these localized pairs, they are not subject to Landau quantization and can act as a reservoir of low energy state necessary for the broadening of the Landau levels. Localized pairs are coupled to the order parameter field $\Phi_j(x)$ through the Andreev reflection mechanism:
\begin{equation}
	\begin{split}
	H_{\text{res}}=&\frac{g}{2}\sum_{j}\int \rd x \left[ \Phi^{\dagger}_j(x)\s^-_j(x)+\s^+_j(x)\Phi_j(x) \right]\\
	&+\frac{1}{2}\sum_j\int \rd x~\tilde{h}_j(x)\s^z_j(x),
	\end{split}
\end{equation}
where the Pauli matrix operator  $\s^+_j(x)$ creates a localized pair at point $x$ within the charged insulator adjacent to the $j$-th superconducting stripe, and the localized level $\tilde{h}_j(x)$ suffers from quenched disorder. To the leading order in the small parameter $g$, the effect of the localized pairs can be represented by a local potential:
\begin{equation}
\label{eq:localp}
	\begin{split}
		& S_{\text{res}}^{\text{eff}}=-\sum_{\omega,j}\int \rd x~ \Phi^{\dagger}_j(\omega,x)V(\omega,x)\Phi_j(\omega,x),\\
		& V(\ri\omega_n,x)=-\frac{g^2}{4}\la \s^-\s^+\ra =g^2 \frac{\tanh(\tilde{h}_j(x)/2T)}{\ri\omega_n -\tilde{h}_j(x)},
	\end{split}
\end{equation}
where the random variable $\tilde{h}_j(x)$ is locally correlated in real space:
\begin{equation}
\label{eq:quench}
	\overline{\tilde{h}_j(x)\tilde{h}_{j'}(x')}=\Delta\delta(x-x')\delta_{jj'}.
\end{equation}
As is shown in Appendix \ref{app:beyond}, the disorder-averaged conductivity receives higher order in $g^2$ contributions from the vertex corrections. As a result, to the leading order in $g^2$, it suffices to use the Drude approximation, where we can average the Green's function individually.

Now we calculate the disorder-averaged Green's function. The leading order contribution from the local potential to the RPA self energy is
\begin{equation}
	\Sigma(\ri\omega_n)=\overline{V(\ri\omega_n,x)}.
\end{equation}
After analytic continuation, its imaginary part is
\begin{equation}
	\Im m \Sigma^R(\omega)=-\ri c\pi g^2 K(\omega), ~~~K(\omega) \propto \tanh (\omega / 2 T),
\end{equation}
where $c$ is the concentration of localized pairs, provided the distribution of disorder is wide enough to encompass the frequency $\omega$. As a result, to the leading order in $g^2$, Eq. (\ref{eq:GG}) is modified as
\begin{equation}
\label{eq:IM}
	\begin{split}
	\bigg{\{}\chi^{-1}_P(\omega,q)&-Je^{\ri pa_0}\delta_{q,q'+h}-Je^{-\ri pa_0}\delta_{q,q'-h}\\
	&+\ri c \pi g^2 K(\omega)\bigg{\}}\overline{G}(\omega,q',q'';p)=\delta_{q,q''},
	\end{split}
\end{equation}
where we have ignored the effect of $\Re e\Sigma^R$, which can be incorporated into renormalization of the Josephson coupling constant. The resulting disorder-averaged Green's function for $h>h_c$ still has the form in Eq. (\ref{eq:disorderfree}), but the function $\chi_k(\omega)$ is modified by the extra imaginary term in Eq. (\ref{eq:IM}). For example, the function $\chi_0(\omega)$ in Eq. (\ref{eq:chi0}) is changed to
\be
\small
\label{eq:IM2}
  \overline{\chi_0}=\frac{1}{\left[(h/h_c)^{\frac{(2-2d)}{(2-d)}} -1\right] - \beta_0\omega^2  - \ri \alpha_0|\omega|^{3-2d}\operatorname{sgn}\omega-\ri\gamma K(\omega)},
\ee
where $\gamma=c\pi g^2 /(2|J|)$. Notably, the disorder-averaged Green's function still respects particle-hole symmetry. 

In the Drude approximation, the disorder-averaged conductivity receives main contributions from products of disorder-averaged Green's functions:
\begin{equation}
	\overline{\Im m\chi_{k_1}(\xi)\Im m\chi_{k_2}(\xi')}\approx \overline{\Im m\chi_{k_1}(\xi)}\cdot\overline{\Im m\chi_{k_2}(\xi')}.
\end{equation}
Then the extra imaginary term in Eq. (\ref{eq:IM}) introduces another contribution to the longitudinal conductivity $\sigma_{xx}$ along the stripe direction in Eq. (\ref{eq:conxx}):
\begin{equation}
\label{eq:Drudepeak}
	\begin{split}
	&\sigma_{xx}(\omega \ll \omega_0)\sim \gamma^2\frac{ \sinh(\omega/2T)}{\omega}\\
	& \times\int \rd y\frac{K(y)K(y+\omega)}{\sinh(y/2T)\sinh\big{(}(y+\omega)/2T\big{)}}+O\left(T^{8-4d}\right)\\
	 &\sim \sigma_0 + O\left(T^{8-4d}\right),
	\end{split}
\end{equation}
where $\omega_0$ corresponds to the lowest Landau level and $\sigma_0 \sim \gamma^2\propto g^4$ is a constant. Consequently, the leading term for $\sigma_{xx}(\omega\to 0)$ is finite even at $T\to 0$, resulting in a Bose metal. As we can see, this anomalous metal does not have a Drude peak. The same analysis also applies to the calculation of the longitudinal conductivity $\sigma_{yy}(\omega\to 0)$ transverse to the stripe direction, while the Hall conductivity $\sigma_{xy}(\omega\to 0)$ still vanishes due to the symmetry argument stated below Eq. (\ref{eq:Hall}). It should be noted, however, that in contrast to the experiment which shows that at high fields the sheet conductance approaches the universal value $\sim 2e^2/2\pi\hbar$, the mechanism we discuss does not lead to such a saturation.

 As far as AC conductivity is concerned, we will be interested in the cyclotron resonance peak. Assuming that the transition frequency $\omega_{k} - \omega_0 \gg T$, from (\ref{eq:F12}) we obtain for $\omega$ near the cyclotron resonance that
 \begin{equation}
 	\begin{split}
 		F_{0,k}(\omega) & \sim \frac{\sinh(\omega/2T)}{\omega}\int_{-\infty}^{\infty} \frac{\rd y}{\cosh(y/2T)\cosh\big{(}(y +\omega)/2T\big{)}} \\
		& \quad \times\frac{\gamma}{\left[\left(y^2 - \omega_0^2\right)^2 + \gamma^2\tanh^2(y/2T)\right]} \\
		& \quad \times\frac{\gamma}{\left[\left[(y+\omega)^2-\omega_k^2\right]^2 +\gamma^2\tanh^2\big{(}(y+\omega)/2T\big{)}\right]} \\
		\xrightarrow{T\to 0} & ~\frac{1}{\omega} \int_{-\omega}^0\rd y \frac{\gamma}{\left(y^2 - \omega_0^2\right)^2 + \gamma^2} \frac{\gamma}{\left[(y+\omega)^2-\omega_k^2\right]^2 +\gamma^2}.
 	\end{split}
 \end{equation}
Since the integrand contains no divergence in the limits of integration, this integral is nonsingular. Therefore, there is no cyclotron resonance at $T \rightarrow 0$. It is easy to see that the absence of the cyclotron resonance at $T\rightarrow 0$ is related to the absence of the Fermi sea for particles with Bose statistics.
Hence for temperatures much smaller than the energy of the first peak, the spectral weight of the cyclotron resonance is small $\sim T$. This is consistent with the experiment \cite{PhysRevLett.120.167002}. It would be interesting to measure the terahertz conductivity for the stripe-ordered LBCO of high disorder level, where the sheet resistance $R_{\square} \gtrsim 6000 \Omega$, with a magnetic field well above 15T. At low temperatures around several to ten Kelvins, our theory suggests that the spectral weight of the cyclotron resonance is negligible. While for higher temperatures, a linear in $T$ dependence starts to emerge. This is in contrast to the more slow development of the Drude peak at higher temperatures as predicted in Eq. (\ref{eq:Drudepeak}).

\section{Discussion and Conclusions}
\label{sec:discuss}

In this paper, we discussed a simple theory of superconducting stripes separated by insulating regions and coupled by Josephson tunneling. Our results suggest that in a sufficiently strong magnetic field the excitation spectrum of the preformed bosonic pairs consists of Landau levels. The corresponding excitations have a finite lifetime due to the interactions, disorder, and temperature effects. In a clean system, the broadening is modest, leading to a power-law temperature dependence of the longitudinal conductivity, which is insufficient to keep the sample metallic above the critical magnetic field when superconductivity is lost. Thus, in a magnetic field, the clean system undergoes a transition from a 2D superconductor to a weak insulator.

 One way to compromise on the observations of the anomalous metal in recent experiments of the stripe-ordered LBCO \cite{tranquada} is to invoke quasiparticles as was suggested in Tsvelik's earlier paper \cite{tsvelik19}. The role of quasiparticles has been discussed in the context of thin films and proximity Josephson junction arrays \cite{Figa1,Figa2,Figa3}, and there are theories where PDW coexists with fermionic quasiparticles \cite{tsvelik16,tsvelik19,lee}. Some of them \cite{tsvelik16,tsvelik19} respect the particle-hole symmetry, while others \cite{lee} do not. 
  
It is then legitimate to ask whether there is any possibility to have a situation where all low energy excitations remain bosonic, yet we still have a metallic state. We have proposed such a possibility in this paper. The essence of our proposal is that strong disorder leads to the creation of localized pairs or, perhaps, small superconducting grains. Due to their small size, their spectrum does not experience Landau quantization and hence provides a reservoir of low energy states for metallic transport. Our proposal provides a mechanism for the Bose metal other than the exotic one of quantum phase glass \cite{PhysRevLett.89.027001,PhysRevB.73.214507}, wherein the latter the Josephson coupling constant is random in sign. The anomalous metallic state resulting from our proposal possesses some of the features observed in the stripe-ordered LBCO \cite{tranquada} and the indium oxide films \cite{PhysRevLett.120.167002}. It has zero Hall response, no Drude peak, and no cyclotron resonance at low temperatures. It would be interesting to measure terahertz conductivity for LBCO to test the relevance of the theory described in this paper.

\section{Acknowledgements}
We are grateful to Andrey Chubukov, Theirry Giamarchi, and John Tranquada for very valuable discussions. We thank Peter Armitage for attracting our attention to Ref. \cite{PhysRevLett.120.167002}.  This work was supported by the Office of Basic Energy Sciences, Material Sciences and Engineering Division, U.S. Department of Energy (DOE) under Contract No. DE-SC0012704.

\newpage
\appendix
\numberwithin{equation}{section}

\section{Effect of Disorder in Josephson Coupling}
In this appendix, we consider the effect of disorder in Josephson coupling on the simple WT. The quenched disorder in the Josephson coupling between superconducting stripes is assumed to be $\delta$-correlated in space:
\begin{equation}
	J\to J+\delta J_j,~~~\overline{\delta J_j(x)\delta J_{j'}(x')}=\Delta_J\delta(x-x')\delta_{jj'},
\end{equation}
where we also assume that the disorder is weak $\Delta_J\ll J^2$ such that it won't change the sign of $J$. We use the replica trick to calculate the disorder-averaged Green's function:
\begin{equation}
\label{eq:rep}
	\begin{split}
	& \overline{G}(t-t',x,x';j-j')\\ 
	=&-\ri\lim_{n\to 0}\int \prod_{a=1}^n\left[\mathcal{D}\bar{\Phi}_j^{(a)}\mathcal{D}\Phi_j^{(a)}\right]\Phi^{(1)}_j(t,x)\bar{\Phi}^{(1)}_{j'}(t',x')e^{\ri S_R},
	\end{split}
\end{equation}
where the replicated action is
\begin{widetext}
\begin{equation}
\label{eq:rep2}
	\begin{split}
	S_R=& \sum_{\omega,q,a,j}\bar{\Phi}_j^{(a)}(\omega,q)\chi_P^{-1}(\omega,q)\Phi^{(a)}_j(\omega,q) - J\sum_{\omega,a,j}\int \rd x \Big[\bar{\Phi}^{(a)}_j(\omega,x)e^{\ri hx}\Phi^{(a)}_{j+1}(\omega,x) + H.c.\Big]  \\
	&+\frac{\ri}{2}\Delta_J\sum_{\omega_1,\omega_2,a,b,j}\int \rd x\Big[\bar{\Phi}^{(a)}_j(\omega_1,x)e^{\ri hx}\Phi^{(a)}_{j+1}(\omega_1,x) + h.c.\Big]\Big[\bar{\Phi}^{(b)}_j(\omega_2,x)e^{\ri hx}\Phi^{(b)}_{j+1}(\omega_2,x) + H.c.\Big].
	\end{split}
\end{equation}
\end{widetext}
It is more convenient to express it in momentum space:
\begin{widetext}
\begin{equation}
	\begin{split}
	S_R=& \sum_{\omega_1,q_1,p_1,a}\bar{\Phi}^{(a)}(\omega_1,q_1;p_1)\Bigg{\{}\chi^{-1}_P(\omega_1,q_1)\Phi^{(a)}(\omega_1,q_1;p_1)-Je^{\ri p_1a_0}\Phi(\omega_1,q_1-h;p_1)  \\
	& -Je^{-\ri p_1a_0}\Phi(\omega_1,q_1+h;p_1) +\frac{\ri}{2}\Delta_J\sum_{\omega_2,q_2,q_2'}\sum_{p_2,p_2'}\sum_b\bar{\Phi}^{(b)}(\omega_2,q_2;p_2)\Phi^{(b)}(\omega_2,q_2';p_2')  \\
	&\Big{[}\Phi^{(a)}(\omega_1,q_1+q_2-q'_2-2h;p_1+p_2-p'_2)e^{\ri(p_1+p_2)a_0}+\Phi^{(a)}(\omega_1,q_1+q_2-q'_2;p_1+p_2-p'_2)e^{\ri(p_1-p'_2)a_0}  \\
	& \Phi^{(a)}(\omega_1,q_1+q_2-q'_2;p_1+p_2-p'_2)e^{-\ri(p_1-p'_2)a_0}+\Phi^{(a)}(\omega_1,q_1+q_2-q'_2+2h;p_1+p_2-p'_2)e^{-\ri(p_1+p_2)a_0} \Big{]}\Bigg{\}}.
	\end{split}
\end{equation}
\end{widetext}
We calculate the one-loop correction to the diagonal Green's function, assuming that the replica symmetry is unbroken. By taking the limit $n\to 0$, we obtain the equation for the disorder-averaged Green's function:
\begin{widetext}
\begin{equation}
	\begin{split}
	& \Bigg{\{}\chi^{-1}_P(\omega_1,q_1)-Je^{\ri p_1a_0}\delta_{q_1,q_1'+h}-Je^{-\ri p_1a_0}\delta_{q_1,q_1'-h}-\frac{1}{2}\Delta_J\sum_{\omega_2,q_2,q_2'}\sum_{p_2}G(\omega_2,q_2',q_2;p_2) \\
	& \Big{[}e^{\ri(p_1+p_2)a_0}\delta_{q_1+q_2-q_2',q_1'+2h}+e^{\ri(p_1-p_2)a_0}\delta_{q_1+q_2-q_2',q_1'}\\
	& +e^{-\ri(p_1-p_2)a_0}\delta_{q_1+q_2-q'_2,q_1'}+e^{-\ri(p_1+p_2)a_0}\delta_{q_1+q_2-q_2',q_1'-2h}\Big{]}\Bigg{\}}\overline{G}(\omega_1,q_1',q_1'';p_1)=\delta_{q_1,q_1''},
	\end{split}
\end{equation}
\end{widetext}
where $G(\omega_2,q'_2,q_2;p_2)$ is the disorder-free Green's function in Eq. (\ref{eq:disfree}). By performing the summation, we arrive at essentially the same equation as in Eq. (\ref{eq:GG}), only the Josephson coupling constant $J$ is replaced by an effective one:
\begin{equation}
\label{eq:renormJ}
	\begin{split}
	J_{\text{eff}}=&J\left\{1+\frac{\Delta_J}{4J^2}\sum_{k=0}^{\infty}\left[\int\frac{\rd\omega}{2\pi}\chi_k(\omega)\right]\right.\\
	&\times\left.\left[\int\frac{\rd q}{2\pi}\Psi_k(q)\Big{(}\Psi_k(q-h)+\Psi_k(q+h)\Big{)} \right] \right\}.
	\end{split}
\end{equation}
For $h<h_c$, we have $\chi_k(\omega)<0$, so the weak disorder weakens the Josephson coupling. For weak disorder $\Delta_J\ll J^2$, the Josephson coupling remains its sign with a smaller magnitude, so we have a stable superconducting phase, only the effective critical magnetic field is reduced.

We now consider the situation where the disorder is weak to maintain the sign of the Josephson coupling, but yet strong enough to reduce the critical magnetic field greatly. Then we can make the approximation by setting $h=0$ in Eq. (\ref{eq:renormJ}), resulting in an effective Josephson coupling constant as
\be
	J_{\text{eff}}=J\left\{1-\frac{\Delta_J}{4\pi^2J^2}\sum_{k=0}^{\infty} \frac{\arctan\sqrt{\beta_k}}{\sqrt{\beta_k}}\left[\int \rd x\Psi_k^2(x) \right] \right\},
\ee
which determines the effective critical magnetic field $h'_c$ through Eq. (\ref{eq:renormH}) by replacing $J$ with $J_{\text{eff}}$. Then the disorder-averaged Green's function $\overline{G}$ can be obtained from Eq. (\ref{eq:disorderfree}) by replacing $h_c$ with $h'_c$, and the analysis of conductivities above $h'_c$ showed in Sec. \ref{sec:conduct} can be carried through without change. As a result, the weak disorder in Josephson coupling between superconducting stripes only shifts the critical point without changing the nature of the phase transition, so it still results in an insulator above the transition point.

The above conclusion can be understood using a coarse-graining picture. In presence of disorder, we will have regions where the Josephson coupling between stripes is strong as well as regions where it is weak. The effective critical magnetic field is determined by the Josephson coupling constant via Eq. (\ref{eq:renormH}). For two stripes where the Josephson coupling between them is strong such that the effective critical magnetic field exceeds the applied magnetic field $h_c>h$, we have superconducting correlation across them and we can fuse them into a single stripe. After one step of such coarse-graining, we end up with exactly the same WT defined in Eq. (\ref{eq:wire}), only with a smaller Josephson coupling constant between the effective stripes. If this coarse-graining procedure can be carried on until we have a single stripe, we are in the superconducting phase. If otherwise, we obtain WT with an effective critical magnetic field $h_c<h$, then the calculation of the conductivities in Sec. \ref{sec:conduct} tells us that we are in the insulating phase. As a result, a Bose metal cannot arise.

\section{Correction Beyond Drude Approximation}
\label{app:beyond}
In this appendix, following Sec. \ref{sec:res}, we show that the correction to the disorder-averaged conductivity beyond the Drude approximation is of higher order in $g^2$. This can be conveniently formulated using the diagrammatic technique. We consider the longitudinal conductivity $\sigma_{xx}$ as an example. From the Kubo formula in Eq. (\ref{eq:Kubo}) we obtain
\begin{equation}
\label{eq:appKubo}
	\begin{split}
		&\sigma_{xx}(\omega)=\frac{1}{4 J^{2} \pi} \int \mathrm{d} y \frac{\operatorname{coth}(y / 2 T)-\operatorname{coth}[(y+\omega) / 2 T]}{\omega}\times\\
		&~~~\int dqdq'qq'\Im m G^R(y,q,q';0)\Im m G^R(y+\omega,q',q;0),
	\end{split}
\end{equation}
where the products of the imaginary part of the Green's functions can be rewritten using both the retarded and advanced Green's functions:
\begin{equation}
	\begin{split}
		\Im m G^R\Im m G^R & = -\frac{1}{4}\left(G^R-G^A\right)\left(G^R-G^A\right)\\
		& =-\frac{1}{2}\Re e\left(G^RG^R-G^RG^A\right)
	\end{split}
\end{equation}
The Drude approximation for the disorder-averaged conductivity replaces the average over products of Green's functions by products of disorder-averaged individual Green's functions:
\begin{equation}
	\overline{G^{R}G^{R}}\approx \overline{G^{R}}\cdot\overline{G^{R}}, ~~~\overline{G^{R}G^{A}}\approx \overline{G^{R}}\cdot\overline{G^{A}}.
\end{equation}
\begin{figure}[htp!]
	\includegraphics[scale=0.09]{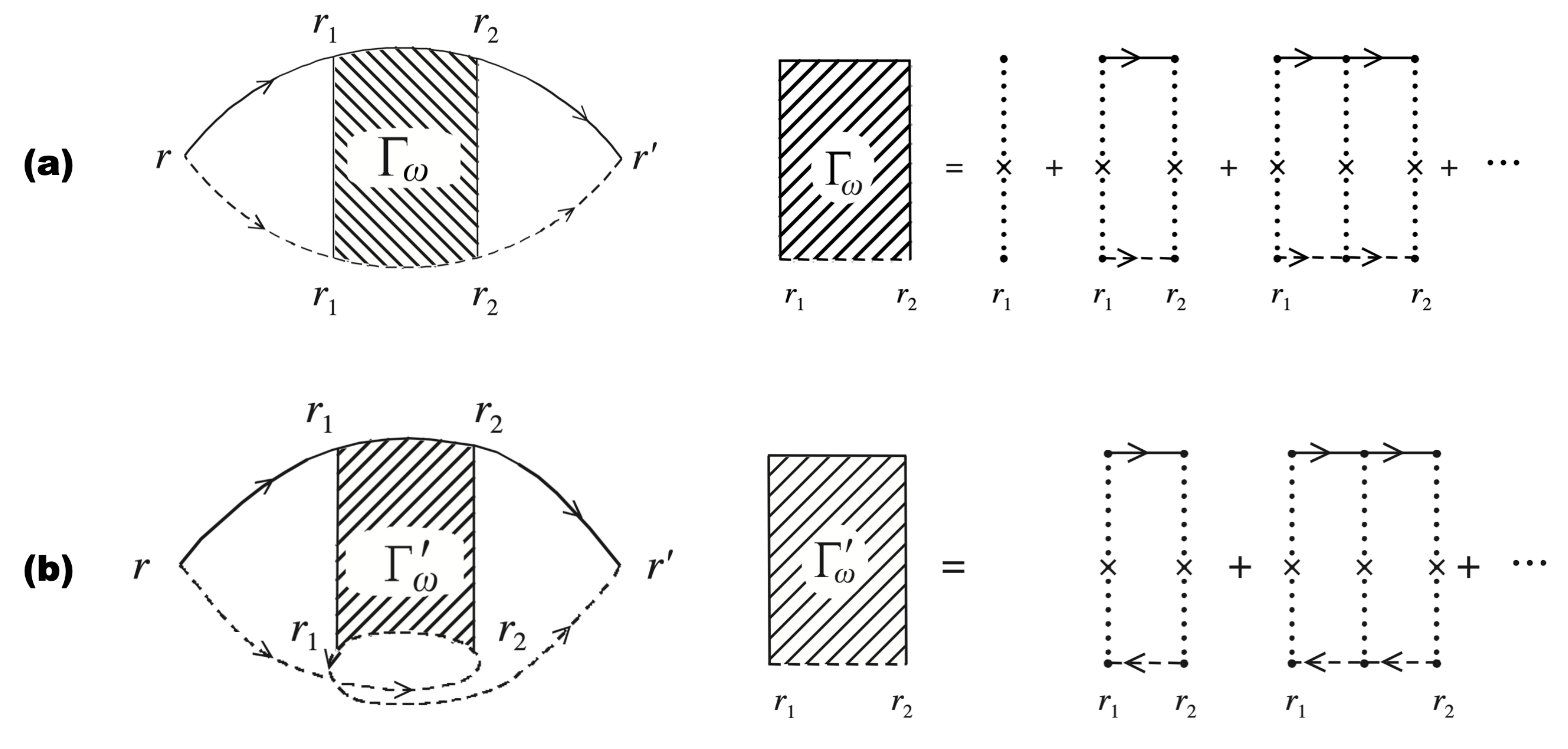}
	\captionsetup{justification=raggedright}
	\caption{Vertex corrections to $\overline{G^RG^A}$ beyond the Drude approximation in real space representation: (a) the diffuson (b) the cooperon. In the figure, $r=(x,j)$ and the cross represents the local potential $V$. The solid and dashed lines represent the retarded Green's function $G^R$ and advanced Green's function $G^A$, respectively.}
	\label{fig:DC}
\end{figure}
We have seen in Sec. \ref{sec:res} that at low temperatures this results in the conductivity $\sigma(\omega\to 0)\propto\gamma^2\propto g^4$. The corrections beyond the Drude approximation come from multiple scatterings on the same localized pair across different propagators, known as the vertex corrections. Two famous examples for $\overline{G^RG^A}$ are the diffuson and the cooperon shown in Fig. \ref{fig:DC}. Besides, in our situation there are also similar contributions for $\overline{G^RG^R}$. These vertex corrections contain the basic element:
\begin{equation}
	\overline{V(\xi,x)V(\xi',x')}=\overline{V(\xi,x)V(\xi',x)}\delta(x-x'),
\end{equation}
which is represented by a dotted line with a cross in Fig. \ref{fig:DC}. When substituting it into the integrals of Eq. (\ref{eq:appKubo}), we just take $\xi=y,\xi'=y+\omega$. The local potential $V(\xi,x)$ is defined in Eq. (\ref{eq:localp}), and the analytic continuation from $\ri\xi_n$ to $\xi\pm \ri 0$ obeys the following rule: if we are dealing with $\overline{G^RG^R}$, then both frequencies are analytically continued as $\ri\xi_n\to \xi+\ri0,\ri\xi'_n\to \xi'+\ri 0$; if we are dealing with $\overline{G^RG^A}$, then the two frequencies are analytically continued differently $\ri \xi_n\to \xi+\ri 0, \ri\xi'_n\to \xi'-\ri 0$.

Given the random distribution characterized by Eq. (\ref{eq:quench}), we obtain the following expressions for $f(\xi,\xi')\equiv \overline{V(\xi,x)V(\xi',x)}$:
\begin{equation}
\label{eq:frrra}
	\begin{split}
		& f^{RR}(\xi,\xi')=\Re ef-\ri c\pi g^4\frac{\tanh^2(\xi/2T)-\tanh^2(\xi'/2T)}{\xi'-\xi},\\
		& f^{RA}(\xi,\xi')=\Re ef-\ri c\pi g^4\frac{\tanh^2(\xi/2T)+\tanh^2(\xi'/2T)}{\xi'-\xi},
	\end{split}
\end{equation}
and $\Re e f$ is the principal integral
\begin{equation}
	\Re e f=g^4~\int_{\text{P.I.}} \rd\tilde{h} P(\tilde{h})\frac{\tanh^2(\tilde{h}/2T)}{(\xi-h)(\xi'-h)},
\end{equation}
where $P(\tilde{h})$ is the distribution of the random variable $\tilde{h}_j(x)$. The corresponding vertex corrections have the important feature that the momentum integration can still be factored out, similar to that in Eq. (\ref{eq:sigmaxx}), such that we can only focus on the frequency integration, which we denote as $\Delta F_{k_1,k_2;k_1',k_2'}(\omega)$. The leading order term of the diffuson-type correction in it is
\begin{figure}
	\includegraphics[scale=0.13]{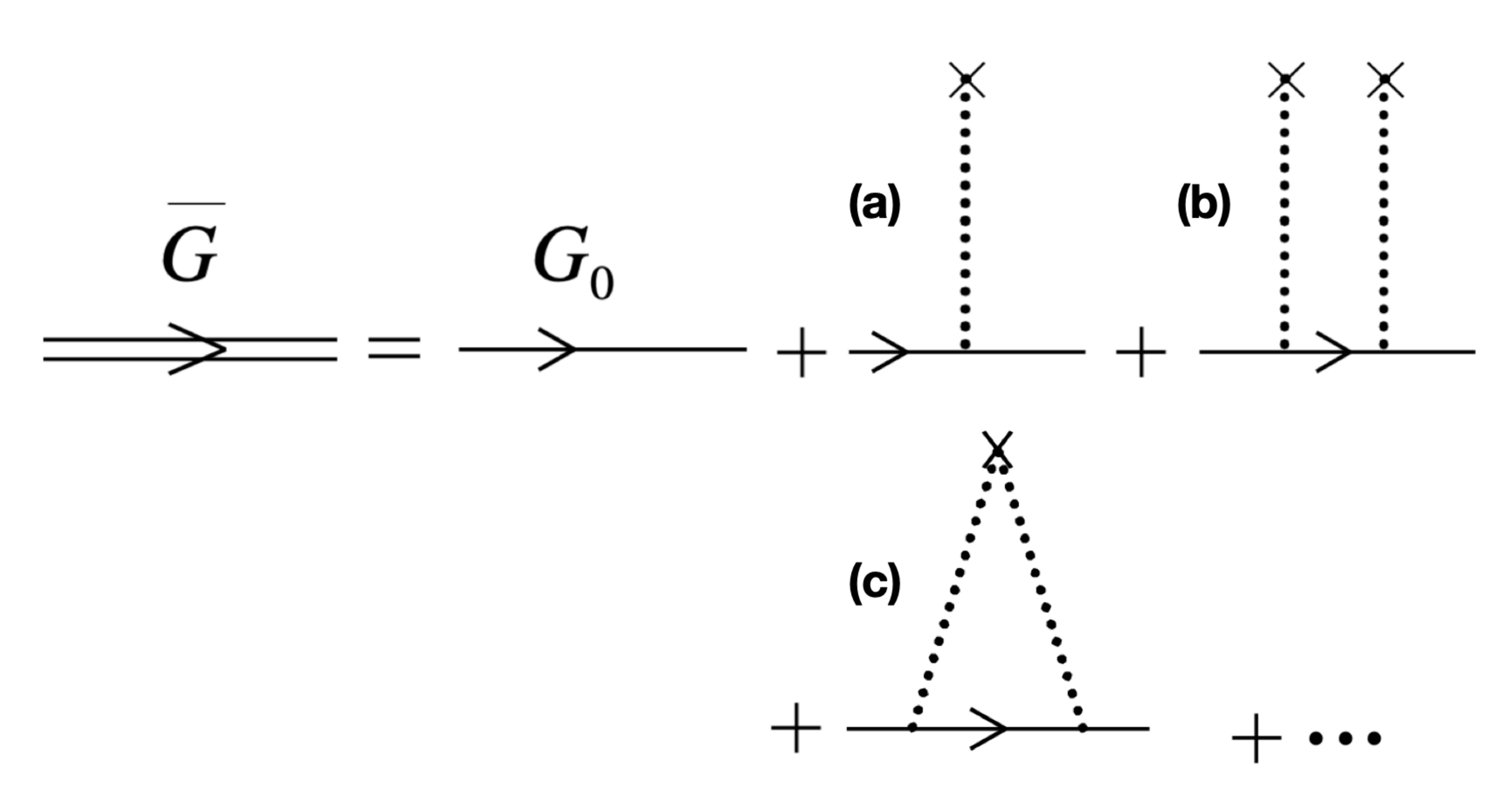}
	\captionsetup{justification=raggedright}
	\caption{Disorder-averaged individual Green's function. In Sec. \ref{sec:res}, we only considered the leading order correction shown in (a). The next order correction includes two terms, shown in (b) and (c).}
	\label{fig:G}
\end{figure}
\begin{widetext}
\begin{equation}
\label{eq:RAd}
	\begin{split}
		& \Delta F_{k_1,k_2;k_1',k'_2}(\omega)=\frac{1}{8J^2\pi} \int \mathrm{d} y \frac{\operatorname{coth}(y / 2 T)-\operatorname{coth}[(y+\omega) / 2 T]}{\omega}\mathcal{D}(y,y+\omega),\\
		& \mathcal{D}(\xi,\xi')= \Re e\Big{[} -P_{k_1,k_2}^{RR}(\xi,\xi')f^{RR}(\xi,\xi')P_{k_1',k_2'}^{RR}(\xi,\xi')+P_{k_1,k_2}^{RA}(\xi,\xi')f^{RA}(\xi,\xi')P_{k_1',k_2'}^{RA}(\xi,\xi')\Big{]},
	\end{split}
\end{equation}
\end{widetext}
where $P_{k_1,k_2}^{RR}(\xi,\xi'),P_{k_1,k_2}^{RA}(\xi,\xi')$ are products of disorder-averaged Green's functions:
\begin{equation}
	\begin{split}
		& P_{k_1,k_2}^{RR}(\xi,\xi')\equiv \overline{\chi^R_{k_1}(\xi)}\cdot \overline{\chi^R_{k_2}(\xi')},\\
		& P_{k_1,k_2}^{RA}(\xi,\xi')\equiv \overline{\chi^R_{k_1}(\xi)}\cdot \overline{\chi^A_{k_2}(\xi')}.
	\end{split}
\end{equation}
Using Eq. (\ref{eq:IM}) and Eq. (\ref{eq:IM2}) for the disorder-averaged Green's function, we can evaluate Eq. (\ref{eq:RAd}), the leading-order diffuson-type correction to the disorder-averaged conductivity as:
\begin{equation}
	\begin{split}
		\sigma_{\mathcal{D}}&(\omega\to 0)\sim~C_1g^4\int \rd y \frac{y^{8-4d}}{2T\sinh^2(y/2T)}\frac{1}{T}\\
		&+C_2g^4\int \rd y\frac{|y|^{3-2d}\operatorname{sgn}y}{2T\sinh^2(y/2T)}\frac{\tanh(y/2T)}{2T\cosh^2(y/2T)}\\
		&+C_3g^4\int\rd y\frac{(3-2d)|y|^{2-2d}\tanh^2(y/2T)}{2T\sinh^2(y/2T)}+O(g^6),
	\end{split}
\end{equation}
where $C_1,C_2,C_3$ are some constants. At low temperatures, we then obtain
\begin{equation}
\label{eq:diffuson}
	\sigma_{\mathcal{D}}(\omega\to0)\sim g^4(T/\epsilon_0)^{2-2d}+O(g^6).
\end{equation}
In the zero temperature limit $T\to 0$, this result is of higher order in $g^2$, compared with the result in the Drude approximation. The same line of calculations can be directly applied to the cooperon-type correction. The leading order term then contains two factors of $f(\xi,\xi')$, as can be seen from Fig. \ref{fig:DC}. Thus, the leading order term of the cooperon-type correction is already of order $g^8$.

There are two divergences that one must be careful with, though. One is due to the $1/\omega$ dependence in the imaginary part of $f^{RA}(y,y+\omega)$, and the other is due to the $1/T$ dependence of all the $f(\xi,\xi')$. In the limit $\omega\to 0,T\to 0$, both of them will lead to divergence of the conductivity in higher order corrections beyond Eq. (\ref{eq:diffuson}), which must be cancelled order by order. The former divergence only happens in diagrams containing no less than two factors of $f^{RA}(y,y+\omega)$, and they will be cancelled off between the diffuson and the cooperon. The latter divergence is cancelled off by the same order correction in the Drude approximation, of which one example is shown in Fig. \ref{fig:G}. There, the next order correction to the disorder-averaged Green's function contains a term (c), which contains the same $1/T$ divergence as that in the next order vertex correction.

In summary, all the vertex corrections are regular and smaller than the leading Drude approximation by factors of $g^2$, so for the disorder-averaged conductivity, the vertex corrections are negligible and it suffices to use the Drude approximation for our model.

\bibliography{Bosepaper.bib}

\end{document}